\documentstyle[preprint,revtex,eqsecnum]{aps}
\tightenlines
\begin{document}
\draft
\preprint{MPI-Ph/92-68}
\begin{title}
Neutrino Flavor Conversion in a Supernova Core
\end{title}
\author{G. Raffelt and G. Sigl}
\begin{instit}
Max-Planck-Institut f\"ur Physik, Postfach 401212, 8000 M\"unchen 40,
Germany
\end{instit}
\begin{abstract}
If $\nu_\mu$ or $\nu_\tau$ mix with $\nu_e$, neutrino oscillations and
collisions in a supernova (SN) core allow these flavors effectively to
participate in $\beta$ equilibrium and thus to obtain a large chemical
potential. If a sterile species mixes with $\nu_e$, these effects
lead to an anomalous loss of energy and lepton number. We study flavor
conversion in a SN core on the basis of a new kinetic equation which
rigorously includes neutrino interference and degeneneracy effects.
Our discussion serves as an example and illustration of the properties
of this ``non-abelian Boltzmann equation''.
\end{abstract}


\section{Introduction}
If neutrinos have masses and if the mass eigenstates do not coincide
with the weak interaction (or flavor) eigenstates, this ``mixing''
leads to oscillations and hence to transitions between different
neutrino flavors: the individual flavor lepton numbers are not
conserved. Therefore, in a thermalized neutrino ensemble, such as in a
supernova (SN) core during and immediately after collapse where
neutrinos are trapped, all mixed flavors will be characterized by the
same chemical potential. Moreover, because the concentration of
electron lepton number is initially rather large, the $\nu_e$ form a
highly degenerate Fermi sea. The other flavors $\nu_\mu$ and
$\nu_\tau$ are characterized by a thermal distribution at zero
chemical potential unless they mix with $\nu_e$ in which case they
would achieve the same large chemical potential.
In a SN core heat and lepton number are transported mostly by
neutrinos; the efficiency of these processes depends crucially on the
degree of neutrino degeneracy for each flavor.
Therefore, it is of great interest to determine the time it takes a
non-$\nu_e$ flavor to equilibrate with $\nu_e$ under the assumption of
mixing \cite{Maalampi,Turner}.

If $\nu_e$ mixes with a sterile neutrino species, conversion into this
inert state leads to the loss of energy and lepton number from the
inner core of a SN. The observed SN 1987A neutrino signal may thus be
used to constrain the allowed range of masses and mixing angles
\cite{Kainulainen}.

Flavor conversion occurs by the combined action of two effects:
Oscillations between neutrino flavors, and collisions with a
``background medium'' which break the coherence between the flavor
components of a mixed state and thus interrupt the evolution of the
phases. One may easily estimate the time scale for achieving chemical
equilibrium to be
\begin{equation}
\tau^{-1}\approx \sin^2(2\Theta)\,D
\label{Xa}
\end{equation}
where $\Theta$ is the neutrino mixing angle in the medium and $D$ is
an effective ``interruption rate'' for the coherent phase evolution,
to be identified with a certain average of the weak scattering rates
involving neutrinos \cite{Stodolsky,Flaig}. Previous discussions
\cite{Maalampi,Turner,Kainulainen} relied essentially on this
reasoning.

Eq.~(\ref{Xa}) is based on a single-particle wave function picture of
neutrino oscillations and thus it is applicable if effects non-linear
in the neutrino density matrices can be ignored. Therefore, this
simple picture does not allow one to include rigorously the effect of
Pauli blocking of neutrino phase space, an effect undoubtedly
important in a SN core where the $\nu_e$ Fermi sea is highly
degenerate. Therefore, we presently revisit the problem of neutrino
flavor conversion on the basis of a newly derived kinetic equation
\cite{RaffeltSS} which rigorously includes neutrino degeneracy
effects. Our discussion differs in several important aspects from
previous works \cite{Maalampi,Turner,Kainulainen}, but we agree with
the overall picture and the numerical estimates to within factors of
order unity.

More importantly, our work serves as a detailed and explicit example
for the use of the ``non-abelian Boltzmann equation'' that was derived
in Ref.~\cite{RaffeltSS} and that we have extended to include neutrino
absorption and production by the background medium. In this equation
the oscillation aspects of neutrino propagation, which are usually
treated as a 2-level Schr\"odinger problem, were unified with its
kinetic aspects. Our present application is a particularly simple and
transparent case because we may use the ``weak damping'' or ``fast
oscillation'' approximation where the neutrinos can be treated as
``propagation eigenstates'' between collisions. We thus obtain a
closed set of differential equations involving only the usual neutrino
occupation numbers for the two mixing flavors.

In Sect.~II we proceed by casting the ``non-abelian Boltzmann
equation'' of Ref.~\cite{RaffeltSS} into a form appropriate for the
``weak damping limit''. In Sect.~III we study flavor conversion
between $\nu_e$ and $\nu_\mu$ or $\nu_\tau$, and we discuss the
emission of sterile neutrinos from a SN core.  In Sect.~IV we
summarize our findings. In Appendix~A we discuss the special case of
two mixing flavors with identical scattering amplitudes on a given
target species and stress that flavor conversion does occur, contrary
to statements found in the literature. In Appendix~B we derive the
neutrino scattering rates needed in Sect.~III.

\newpage


\section{Kinetic Equation for Mixed Neutrinos}
\subsection{General Result}
The usual objects of study for a kinetic description of a neutrino
ensemble interacting with a thermal heat bath are the occupation
numbers $f_{\bf p}=\langle b^\dagger({\bf p})b({\bf p})\rangle$ of a
mode $\bf p$ of the neutrino field, where $b^\dagger({\bf p})$ is a
creation operator for a neutrino in this mode and the expectation value
refers to the state of the ensemble, not to the vacuum. The physical
situation of interest is one with highly degenerate neutrino Fermi seas
so that we may ignore the presence of anti-neutrinos which otherwise
would have to be included in a kinetic description of the system.

The effect of neutrino mixing between $n$ flavors can be rigorously
included if we replace $f_{\bf p}$ by a $n{\times}n$ ``matrix of
densities'' $\rho_{ij}({\bf p})\equiv\langle b^\dagger_j({\bf p})
b_i({\bf p})\rangle$, $i,j=1,\ldots,n$, where $b^\dagger_i({\bf p})$ is
a creation operator for a neutrino of flavor $i$ and momentum $\bf p$
\cite{RaffeltSS,Dolgov}. The diagonal elements of $\rho_{\bf
p}\equiv\rho({\bf p})$ are the usual occupation numbers $f^i_{\bf p}$
for each flavor $i=1,\ldots,n$ while the off-diagonal terms contain
more subtle phase information.

Earlier, Raffelt, Sigl, and Stodolsky \cite{RaffeltSS} have derived a
kinetic equation for the matrices $\rho_{\bf p}$ of a homogeneous and
isotropic ensemble under the assumption that all neutrino flavors $i$
scatter on a given species of targets in the same way apart from an
overall amplitude factor $g_i$. If there are several species $a$ of
targets present, and if the corresponding quantum fields are
uncorrelated, the collision integral is
[notation $d{\bf p}\equiv d^3{\bf p}/(2\pi)^3$]
\begin{eqnarray}
\dot\rho_{\bf p}=i[\rho_{\bf p},\Omega_{\bf p}]
+\sum_a \int d{\bf p'}&&\biggl[
{\cal W}^a_{{\bf p'}{\bf p}}\,
\frac{G^a\rho_{\bf p'} G^a(1-\rho_{\bf p})+{\rm h.c.}}{2}\nonumber\\
&&-{\cal W}^a_{{\bf p}{\bf p'}}\,
\frac{\rho_{\bf p} G^a(1-\rho_{\bf p'})G^a+{\rm h.c.}}{2}\biggr],
\label{Xb}
\end{eqnarray}
where the $n{\times}n$ matrix
$G^a\equiv{\rm diag}(g_1^a,\ldots,g_n^a)$ in the neutrino interaction
basis. Moreover, $(g_i^a)^2{\cal W}^a_{{\bf p}{\bf p'}}\,$ is the
transition probability for a neutrino $i$ with momentum $\bf p$ into a
state $\bf p'$ due to the interactions with the medium species $a$.
${\cal W}^a$ depends on the neutrino interactions with the medium
constituents and on the intrinsic interactions of the medium.  Finally,
$\Omega_{\bf p}$ is a matrix of oscillation frequencies; its
eigenvalues are the energies in the medium of the $n$ neutrino flavors
at momentum $\bf p$.
If there is only one flavor Eq.~(\ref{Xb}) reduces to the usual
Boltzmann collision integral because $\rho_{\bf p}\to f_{\bf p}$,
the oscillation term disappears, and $G^a$ becomes a simple number.

Neutrino self-interactions are not included in this description.
Because in a SN core the neutrino chemical potential is always much
smaller than that of the electrons it is physically justified to
neglect neutrino-neutrino interactions relative to other targets.

Besides neutrino scattering we also need to include their absorption or
production by the medium. Apart from pair-processes involving
anti-neutrinos, which we continue to ignore, this can occur by
charged-current reactions where a $\nu_e$, for example, is transformed
into an $e^-$ or vice versa. We have derived the corresponding term for
our non-abelian Boltzmann equation by following the same line of
reasoning as in Ref.~\cite{RaffeltSS}, except that we replace the
neutral currents involving neutrino field bilinears with a charged
current involving a neutrino and a charged-lepton field.

Our result can be represented in a compact form by means of the
$n{\times}n$ matrices $I_i$ which, in the interaction basis, contain
only zeros except for a 1 on the $i$-th diagonal position. Then we find
\begin{equation}
\dot\rho_{\bf p}=\ldots+ \sum_{i=1}^n
\bigl(I_i-{\textstyle{1\over2}}\{I_i,\rho_{\bf p}\}\bigr)\,
{\cal P}_{\bf p}^i
-{\textstyle{1\over2}}\{I_i,\rho_{\bf p}\}\,{\cal A}_{\bf p}^i,
\label{Xc}
\end{equation}
where ``$\ldots$'' stands for the terms already present in
Eq.~(\ref{Xb}) and $\{\,{\cdot}\,,\,{\cdot}\,\}$ is an
anti-commutator. As before, we assume that the effect of the neutrino
interactions on the medium can be neglected so that
${\cal P}_{\bf p}^i$, the production rate of a neutrino flavor $i$ with
momentum $\bf p$, and ${\cal A}_{\bf p}^i$, the absorption rate of a
neutrino $i$ with momentum $\bf p$, are given in terms of the
equilibrium properties of the medium.
For a single flavor the r.h.s.\ of Eq.~(\ref{Xc}) takes on the familiar
form of the difference between a rate-of-gain and -loss term,
$(1-f_{\bf p}){\cal P}_{\bf p}-f_{\bf p}{\cal A}_{\bf p}$,
including the usual Pauli blocking factor.

We will always assume isotropy of the medium and of the neutrino
ensemble so that all functions of $\bf p$ really depend only on
$E=\vert{\bf p}\vert$. Therefore, we define the angular average of the
transition rate
\begin{equation}
{\cal W}^a_{EE'}\equiv\frac{1}{4\pi}\int
{\cal W}^a_{\bf pp'} d\Omega
\label{Zg}
\end{equation}
where $\int d\Omega$ is an angular integration about the relative angle
between $\bf p$ and $\bf p'$. If we substitute this expression
everywhere for ${\cal W}_{\bf pp'}$ we may also substitute
$d{\bf p}=d^3{\bf p}/(2\pi)^3\to E^2dE/2\pi^2$. Throughout most of our
discussion we will prefer, however, to keep ${\bf p}$ and ${\bf p'}$ as
our variables.

Equations~(\ref{Xb}) and~(\ref{Xc}) are the ``non-abelian Boltzmann
collision integrals'' which serve as a starting point for a kinetic
treatment of neutrino flavor conversion in a SN core.

\subsection{Two-Flavor Mixing}
In what follows we shall focus on two-flavor mixing, $n=2$. In this
case the oscillation term in Eq.~(\ref{Xb}) can be simplified if we
express all $2{\times}2$ matrices through Pauli matrices. Notably, we
write $\rho_{\bf p}={1\over2}(n_{\bf p}+{\bf P}_{\bf p} \cdot\tau)$ and
$\Omega_{\bf p}=E_{\bf p}+{1\over2}{\bf B}_{\bf p} \cdot\tau$, where
$n_{\bf p}\equiv{\rm Tr}(\rho_{\bf p})$ is the total number of
neutrinos in mode $\bf p$, $E_{\bf p}\equiv{1\over2}{\rm
Tr}(\Omega_{\bf p})$ is the average of the energy eigenvalues of this
mode, and $\tau$ is a three-vector of Pauli matrices. Oscillations
alone then lead to the well-known precession formula
\begin{equation}
\dot{\bf P}_{\bf p}={\bf B}_{\bf p}\times{\bf P}_{\bf p}.
\label{Xe}
\end{equation}
The ``magnetic field'' ${\bf B}_{\bf p}\equiv{\bf B}({\bf p})$ around
which the ``polarization vector'' precesses is given by the standard
result
\begin{eqnarray}
B_1({\bf p})&=&\frac{\Delta m^2}{2\vert{\bf p}\vert}\,
\sin 2\Theta_0,
\nonumber\\
B_3({\bf p})&=&\frac{\Delta m^2}{2\vert{\bf p}\vert}\,
\cos 2\Theta_0-
\sqrt2 G_{\rm F} N,
\label{Xf}
\end{eqnarray}
where we have taken $B_2({\bf p})=0$ without loss of generality,
$G_{\rm F}$ is the Fermi constant, $\Theta_0$ is the vacuum mixing
angle, $\Delta m^2$ is the difference of the squared neutrino vacuum
mass eigenvalues, and we have used a relativistic approximation.
Ignoring neutrino self-interactions which would provide only a small
correction in the physical situation of interest we need to use
$N=N_e-N_{\mu,\tau}$ for the mixing between $\nu_e$ and
$\nu_{\mu,\tau}$.  For the mixing of $\nu_{e,\mu,\tau}$ with a sterile
species $\nu_x$ we have $N=N_{e,\mu,\tau}-{1\over2}N_n$.

The angle of the vector ${\bf B}$ against the 3-axis in flavor space is
$2\Theta_{\bf p}$, twice the usual medium mixing angle, which is a
function of the neutrino momentum. It is given by
\begin{equation}
t_{\bf p}\equiv
\tan 2\Theta_{\bf p}=\frac{\sin2\Theta_0}{\cos2\Theta_0 - E/E_r},
\label{Xi}
\end{equation}
where in our relativistic approximation $E=\vert{\bf p}\vert$ and
\begin{equation}
E_r\equiv\frac{\Delta m^2}{2\sqrt2 G_{\rm F} N}
\label{Xii}
\end{equation}
is a ``resonance energy''. For $\Theta_0\ll1$, a limit which we shall
usually take, neutrinos with the energy $E_r$ experience maximum mixing
with $\vert\Theta_{\bf p}\vert=\pi/4$. Besides $t_{\bf p}$ we will also
need
$s_{\bf p}\equiv\sin 2\Theta_{\bf p}$ and
$c_{\bf p}\equiv\cos 2\Theta_{\bf p}$. We have
\begin{equation}
s_{\bf p}^2=
\frac{4\Theta_0^2}{4\Theta_0^2 +(1-E/E_r)^2}
\label{Xiii}
\end{equation}
where, again, we have assumed $\Theta_0\ll 1$.

We will mostly focus on the mixing of $\nu_e$ with some other flavor
and will always assume that only $\nu_e$'s can be produced or absorbed
by the medium, while the other flavors can only scatter.  Therefore,
${\cal P}_{\bf p}$ and ${\cal A}_{\bf p}$ without superscript will
always refer to the production and absorption rate of $\nu_e$.  Of
course, if a degenerate sea of $\nu_\mu$'s is produced by flavor
conversion, the Fermi energy can exceed the muon production threshold,
and in this case charged-current processes involving muons would have
to be included. As an indication that we will exclude this possibility
in our discussion we will use $\nu_e$-$\nu_\tau$-oscillations as our
prime example.

\subsection{Weak Damping Limit}
Even for two-flavor mixing the general form of the collision integral
remains rather complicated. However, for the conditions of a SN core we
may apply a further approximation, the ``weak damping limit''. It is
easy to show that for the physical conditions of a SN core the
oscillation rate for $\rho_{\bf p}$ (the precession rate of ${\bf
P}_{\bf p}$) is much faster than the scattering rate. Therefore, it is
justified to consider density matrices $\overline\rho_{\bf p}$ averaged
over a period of oscillation. While the matrices
$\rho_{\bf p}$ are given by the four real parameters $n_{\bf p}$ and
${\bf P}_{\bf p}$ which are functions of time, the matrices
$\overline\rho_{\bf p}$ require only two such functions, for example
$n_{\bf p}$ and $\vert{\bf P}_{\bf p}\vert$, while the direction of
${\bf P}_{\bf p}$ remains fixed and is identical with that
of~${\bf B}_{\bf p}$.

One way of looking at the weak damping limit is that between
collisions, rather than using weak interaction eigenstates, neutrinos
are best described by ``propagation eigenstates'', i.e., in a basis
where the $\overline\rho_{\bf p}$ are diagonal. In this basis the
matrices of coupling constants $G^a$ are no longer diagonal, whence
flavor conversion is understood as the result of ``flavor-changing
neutral currents''. Of course, ``flavor'' now refers to the propagation
eigenstates.
However, because in general the effective mixing angle is a function of
the neutrino energy one would have to use a different basis for each
momentum state, an approach that complicates rather than simplifies the
equations. Therefore, we will always work in the interaction basis.

In order to derive an equation of motion for $\overline\rho_{\bf p}$ we
evaluate the r.h.s.\ of Eqs.~(\ref{Xb}) and~(\ref{Xc}) in terms of the
$\overline\rho_{\bf p}$.  We expand the result in Pauli matrices,
leading to an expression of the form ${1\over2}(a_{\bf p}+{\bf A}_{\bf
p}\cdot\tau)$.  While in general the vector ${\bf A}_{\bf p}$ is not
parallel to ${\bf B}_{\bf p}$, the assumed fast oscillations average
the perpendicular component of ${\bf A}_{\bf p}$ to zero so that only
the parallel component remains.  Therefore,
\begin{equation}
\dot{\overline \rho}_{\bf p}={\textstyle{1\over2}}
\bigl[a_{\bf p}+(\hat{\bf B}_{\bf p}\cdot{\bf A}_{\bf p})\,
(\hat{\bf B}_{\bf p}\cdot\tau)\bigr],
\label{Xj}
\end{equation}
where $\hat{\bf B}_{\bf p}\equiv{\bf B}_{\bf p}/
\vert{\bf B}_{\bf p}\vert=(s_{\bf p},0,c_{\bf p})$.

For our purposes the most practical quantities to parametrize the
matrices $\overline\rho_{\bf p}$ are their diagonal entries, the
occupation numbers of the two mixing flavors which we take to be
$\nu_e$ and $\nu_\tau$.  It is straightforward to show that
\begin{equation}
\overline\rho_{\bf p}=\pmatrix{f^{e}_{\bf p}&0\cr
                            0&f^{\tau}_{\bf p}\cr}
   +\frac{(f^{e}_{\bf p}-f^{\tau}_{\bf p})t_{\bf p}}{2}\,
   \pmatrix{0&1\cr1&0\cr}
\label{Xh}
\end{equation}
where $f^{e,\tau}_{\bf p}$ are the occupation
numbers\footnote{For the charged leptons, we denote the occupation
numbers by $n^e_{\bf p}$ and $n^\tau_{\bf p}$. The chemical potentials
and number densities are denoted by $\mu_{\nu_{e,\tau}}$ and
$N_{\nu_{e,\tau}}$ for the neutrinos and $\mu_{e,\tau}$ and
$N_{e,\tau}$ for the charged leptons.} of $\nu_{e,\tau}$.

\newpage

{}From Eq.~(\ref{Xj}) we then find explicitly for the ``absorption
terms'' Eq.~(\ref{Xc}) of the collision integral
\begin{eqnarray}
\dot f_{\bf p}^{e}&=&
\Bigl[(1-f_{\bf p}^{e})\,{\cal P}_{\bf p}
-f_{\bf p}^{e}\,{\cal A}_{\bf p}\Bigr]
-\frac{s_{\bf p}^2}{2}\,
\left[\left(1-\frac{f_{\bf p}^{e}+f_{\bf p}^{\tau}}{2}\right)
{\cal P}_{\bf p}
-\frac{f_{\bf p}^{e}+f_{\bf p}^{\tau}}{2}\,{\cal A}_{\bf p}
\right],\nonumber\\
\dot f_{\bf p}^{\tau}&=&
\frac{s_{\bf p}^2}{2}\,
\left[\left(1-\frac{f_{\bf p}^{e}+f_{\bf p}^{\tau}}{2}\right)
{\cal P}_{\bf p}
-\frac{f_{\bf p}^{e}+f_{\bf p}^{\tau}}{2}\,{\cal A}_{\bf p}
\right].
\label{Ya}
\end{eqnarray}
If neither $\nu_e$ nor $\nu_\tau$ are occupied because, for example,
the medium is transparent to neutrinos so that they escape after
production, we have $\dot f_{\bf p}^{\tau}
={1\over2}s^2_{\bf p}{\cal P}_{\bf p}$, i.e., the production rate of
$\nu_\tau$ is that of $\nu_e$ times ${1\over 2}\sin^2(2\Theta)$,
in agreement\footnote{There is a small discrepancy with Turner's
\cite{Turner} discussion who used an average production efficiency of
$\sin^2\Theta$ instead, which for small mixing angles is half of our
result. He argued that, if $\nu_e$ was a superposition
$\cos\Theta\,\nu_1+\sin\Theta\,\nu_2$ of mass-eigenstates, the
production amplitude of $\nu_2$ was $\sin\Theta$ times that of $\nu_e$
and its production probability, therefore, $\sin^2\Theta$ times that of
$\nu_e$.  For small $\Theta$ one has $\nu_\tau\approx\nu_2$, hence for
small $\Theta$ the production probability for $\nu_\tau$ appears to be
$\sin^2\Theta$ times that of $\nu_e$. This line of argument, however,
ignores the interference between $\nu_1$ and $\nu_2$ which leads to the
oscillation phenomena.  In the usual single-particle treatment the
probability for finding a $\nu_\tau$ at a distance $L$ from the
production site of a $\nu_e$ is $\sin^2(2\Theta)\,\sin^2(\pi L/L_{\rm
osc})$ where $L_{\rm osc}$ is the oscillation length. Thus, on average
one expects ${1\over 2}\sin^2(2\Theta)$ neutrinos of the ``wrong''
flavor for each $\nu_e$ produced.} with Ref.~\cite{Kainulainen}.

In a SN core where the interacting neutrino species are trapped
Eq.~(\ref{Ya}) is more complicated as backreaction and Pauli blocking
effects must be included. It becomes simple, again, if $s_{\bf p}^2\ll
1$, because then the $\nu_e$ will reach $\beta$ equilibrium much faster
than the $\nu_\tau$; at this point we may use to lowest order the
detailed-balance condition
\begin{equation}
f^e_{\bf p}({\cal P}_{\bf p}+{\cal A}_{\bf p})={\cal P}_{\bf p}.
\label{Yaa}
\end{equation}
Inserting this into Eq.~(\ref{Ya}) leads to
$\dot f^\tau_{\bf p}={1\over 4} s^2_{\bf p}
[(1-f_{\bf p}^\tau)\,{\cal P}_{\bf p}
-f_{\bf p}^\tau\,{\cal A}_{\bf p}\Bigr]$
so that now the $\nu_\tau$ follow a naive Boltzmann collision equation
with rates of gain and loss given by those of $\nu_e$ times
${1\over4}\sin^2(2\Theta)$. If we consider the emission of a sterile
species $\nu_x$ instead of $\nu_\tau$, they would escape without
building up so that their production rate would be
${1\over4}\sin^2(2\Theta)$ that of $\nu_e$, half as much compared to
the above situation where the $\nu_e$ were also free to
escape\footnote{In Ref.~\cite{Turner} this factor $1/2$ was not
included, precisely compensating the missing factor of 2 mentioned
before. In Ref.~\cite{Kainulainen} this factor was also left out.}.

\newpage

For the scattering terms Eq.~(\ref{Xb}) of the collision integral we
find from Eq.~(\ref{Xj}) explicitly
\begin{eqnarray}
\dot f^\tau_{\bf p}&=&{1\over4}\int d{\bf p'}\biggl\{
\Bigl[{\cal W}_{\bf p'p}f^\tau_{\bf p'}(1-f^\tau_{\bf p})
     -{\cal W}_{\bf pp'}f^\tau_{\bf p}(1-f^\tau_{\bf p'})\Bigr]
\Bigl[(4-s_{\bf p}^2)g_\tau^2 +
      (2-s_{\bf p}^2) t_{\bf p} t_{\bf p'}g_eg_\tau \Bigr]
\nonumber\\
&&\hskip3.16em
+\Bigl[{\cal W}_{\bf p'p}f^\tau_{\bf p'}(1-f^e_{\bf p})
     -{\cal W}_{\bf pp'}f^e_{\bf p}(1-f^\tau_{\bf p'})\Bigr]
\Bigl[-s_{\bf p}^2 g_\tau^2
      -s_{\bf p}^2 t_{\bf p} t_{\bf p'}g_eg_\tau \Bigr]
\nonumber\\
&&\hskip3.16em
+\Bigl[{\cal W}_{\bf p'p}f^e_{\bf p'}(1-f^\tau_{\bf p})
     -{\cal W}_{\bf pp'}f^\tau_{\bf p}(1-f^e_{\bf p'})\Bigr]
\Bigl[s_{\bf p}^2 g_e^2
      -(2-s_{\bf p}^2)t_{\bf p} t_{\bf p'}g_eg_\tau \Bigr]
\nonumber\\
&&\hskip3.16em
+\Bigl[{\cal W}_{\bf p'p}f^e_{\bf p'}(1-f^e_{\bf p})
     -{\cal W}_{\bf pp'}f^e_{\bf p}(1-f^e_{\bf p'})\Bigr]
\Bigl[s_{\bf p}^2 g_e^2
      +s_{\bf p}^2t_{\bf p} t_{\bf p'}g_eg_\tau \Bigr]
\biggr\}.
\label{Yc}
\end{eqnarray}
For notational simplicity we have written
$G^a={\rm diag}(g_e^a,g_\tau^a)$, i.e., the indices $e$ and $\tau$
refer, again, to $\nu_e$ and $\nu_\tau$. Moreover, we have suppressed
the superscript $a$ everywhere; a summation over target species is
understood. The equation of motion for $f^e_{\bf p}$ is the same if we
exchange $e\leftrightarrow\tau$ everywhere. In the absence of mixing we
have $s_{\bf p}=t_{\bf p}=0$, leading to the usual collision integral
for each species separately.
Integrating Eq.~(\ref{Yc}) over $d{\bf p}$ yields
\begin{eqnarray}
\dot N_{\nu_\tau}&=&{1\over4}\int\!d{\bf p}\!\int\!d{\bf p'}\,
\biggl\{{\cal W}_{\bf pp'}(s_{\bf p}^2-s_{\bf p'}^2)
\Bigr[f^\tau_{\bf p}(1-f^\tau_{\bf p'})
(g_\tau^2+t_{\bf p}t_{\bf p'} g_eg_\tau)
\label{Yd}\\
&&\hskip13em
-f^e_{\bf p}(1-f^e_{\bf p'})
(g_e^2+t_{\bf p}t_{\bf p'} g_eg_\tau)\Bigr]
\nonumber\\
&&\hskip5em
+\Bigl[{\cal W}_{\bf p'p}f^e_{\bf p'}(1-f^\tau_{\bf p})
-{\cal W}_{\bf pp'}f^\tau_{\bf p}(1-f^e_{\bf p'})\Bigr]
\nonumber\\
&&\hskip6em
\times
\Bigl[(s_{\bf p}^2g_e^2+s_{\bf p'}^2g_\tau^2
      -2t_{\bf p}t_{\bf p'}g_eg_\tau)
     +(s_{\bf p}^2+s_{\bf p'}^2)\,t_{\bf p}t_{\bf p'}g_eg_\tau
\Bigr]\biggr\}
\nonumber
\end{eqnarray}
for the rate of change of the total $\nu_\tau$ number density due to
elastic scattering.

We will also have occasion to consider the mixing of $\nu_e$ with a
sterile species $\nu_x$ for which $g_x=0$ by definition. This yields
for the scattering part of the collision integral
\begin{equation}
\dot f^x_{\bf p}={s^2_{\bf p}\over4}\,g_e^2
\int\! d{\bf p'}
\Bigl[{\cal W}_{\bf p'p}f^e_{\bf p'}(2-f^e_{\bf p}-f^x_{\bf p})
     -{\cal W}_{\bf pp'}(f^e_{\bf p}+f^x_{\bf p})(1-f^e_{\bf p'})
\Bigr].
\label{Ycc}
\end{equation}
If the $\nu_e$ stay approximately in thermal equilibrium, detailed
balance allows for a further simplification. Together with the
absorption term Eq.~(\ref{Ya}) and with Eq.~(\ref{Yaa}) we find in this
approximation
\begin{equation}
\dot f^x_{\bf p}=
\frac{s_{\bf p}^2}{4}\left\{\Bigl[
(1-f_{\bf p}^x){\cal P}_{\bf p}-f_{\bf p}^x\,{\cal A}_{\bf p}\Bigr]
+g_e^2\int\!d{\bf p'}
\Bigl[{\cal W}_{\bf p'p}f^e_{\bf p'}(1-f^x_{\bf p})
     -{\cal W}_{\bf pp'}f^x_{\bf p}(1-f^e_{\bf p'})
\Bigr]
\right\}.
\label{Yccc}
\end{equation}
If the mixing angle is so small that the $\nu_x$ freely escape we may
set $f^x_{\bf p}=0$ on the r.h.s.\ of this equation and find
\begin{equation}
\dot N_L=\int\!d{\bf p}
\frac{s_{\bf p}^2}{4}\left[{\cal P}_{\bf p} + \int\!d{\bf p'}\,g_e^2
{\cal W}_{\bf p'p}f^e_{\bf p'}
\right]
\label{Ycccc}
\end{equation}
for the emission rate of lepton number per unit volume. For the
energy-loss rate we need to include a factor $E=\vert{\bf p}\vert$ in
the integral.

\subsection{Small Mixing Angle}
In order to treat neutrino flavor conversion in a SN core it is not
necessary to solve Eqs.~(\ref{Ya}) and~(\ref{Yc}) in their full
complexity. In a medium, maximum mixing obtains if the denominator of
Eq.~(\ref{Xi}) vanishes. For such conditions flavor equilibrium occurs
practically on the time scale it takes to achieve $\beta$ equilibrium
which is instantaneous compared with other relevant time scales. In
this case a detailed evaluation of the kinetic equations is
superfluous. Therefore, we may focus on a situation where the neutrino
mixing angle is small, $\Theta_{\bf p}\ll 1$.

For small mixing angles $\beta$ processes as well as elastic neutrino
collisions with medium particles are much faster than the rate of
flavor conversion which is suppressed by a factor $\Theta^2$.
Therefore, both $f^\tau_{\bf p}$ and $f^e_{\bf p}$ will be given to
lowest order by Fermi-Dirac distributions which are characterized by
slowly varying chemical potentials $\mu_{\nu_\tau}(t)$ and
$\mu_{\nu_e}(t)$. The integrated version of Eq.~(\ref{Ya}) with the
detailed-balance condition Eq.~(\ref{Yaa}), together with
Eq.~(\ref{Yd}) then yields to lowest order in $\Theta^2$
\begin{eqnarray}
\dot N_{\nu_\tau}
&=&\int\!d{\bf p}\,
\Theta_{\bf p}^2 \Bigl[(1-f_{\bf p}^{\tau})\,{\cal P}_{\bf p}
-f_{\bf p}^{\tau}{\cal A}_{\bf p}\Bigr]
\nonumber\\
&+&\int\!d{\bf p}\!\int\!d{\bf p'} {\cal W}_{\bf pp'}\biggl\{
(\Theta_{\bf p}^2-\Theta_{\bf p'}^2)
\Bigl[g_\tau^2 f^\tau_{\bf p}(1-f^\tau_{\bf p'})
      -g_e^2 f^e_{\bf p}(1-f^e_{\bf p'})\Bigr]\nonumber\\
&&\hskip7.5em+\,(g_\tau\Theta_{\bf p}-g_e\Theta_{\bf p'})^2
              f^e_{\bf p}(1-f^\tau_{\bf p'})\nonumber\\
&&\hskip7.5em-\,(g_e\Theta_{\bf p}-g_\tau\Theta_{\bf p'})^2
              f^\tau_{\bf p}(1-f^e_{\bf p'})\biggr\},
\label{Ze}
\end{eqnarray}
and a similar equation for $\dot N_{\nu_e}$. Together with the
condition of $\beta$ equilibrium, $\mu_n-\mu_p=\mu_e-\mu_{\nu_e}$, that
of charge neutrality, $N_p=N_e$, and the conservation of the trapped
lepton number, $d(N_e+N_{\nu_e}+N_{\nu_\tau})/dt=0$, these equations
represent differential equations for the chemical potentials
$\mu_{\nu_\tau}(t)$ and $\mu_{\nu_e}(t)$.

Flavor conversion redistributes the lepton number trapped in a SN core
among $e^-$, $\nu_e$, and $\nu_\tau$. However, most of the lepton
number resides in electrons and so, only a small change of the $e^-$
and $\nu_e$ chemical potentials occur. Taking them at their unperturbed
equilibrium values Eq.~(\ref{Ze}) simply is a differential equation for
$\mu_{\nu_\tau}(t)$.

\newpage

\subsection{Very Degenerate Neutrinos}
In a SN core the $\nu_e$ Fermi sea is very degenerate. With regard to
the neutrino distributions we may thus use the approximation $T=0$ so
that neutrino occupation numbers are~1 below their Fermi surface, and~0
above. Also, elastic scatterings on medium particles will always lead
to neutrino down-scattering because the medium can not transfer energy
to the neutrinos: ${\cal W}^a_{EE'}=0$ for $E'>E$.  This implies
immediately that the term proportional to $(\Theta_{\bf p}^2-
\Theta_{\bf p'}^2)$ in Eq.~(\ref{Ze}) vanishes.  Moreover, because
$\mu_{\nu_e}>\mu_{\nu_\tau}$ the term proportional to $f^\tau_{\bf
p}(1-f^e_{\bf p'})$ also vanishes. If we further observe that
Eq.~(\ref{Yaa}) implies\footnote{In Appendix~B we will calculate
${\cal P}_E$ from the process $e+p\to n+\nu_e$, taking the electrons to
be highly degenerate, and the nucleons to be non-degenerate. In this
limit one could be tempted to extend the integration up to $\mu_e$,
i.e., to use ${\cal P}_E=0$ for $E>\mu_e$ rather than for
$E>\mu_{\nu_e}$, as was done, for example, in Ref.~\cite{Turner}. This
procedure, however, would violate the detailed-balance requirement
Eq.~(\ref{Yaa}) and thus, the approximations would not be
self-consistent. In the final answer the upper limit of integration
appears in a high power, the fifth power in the ``vacuum limit'' (see
Sect.~III). Therefore, even though $\mu_{\nu_e}$ is not much smaller
than $\mu_e$ in a SN core, $\mu_{\nu_e}^5$ {\it is\/} much smaller than
$\mu_e^5$, perhaps by as much as a factor of 10.}
${\cal P}_E=0$ for $E>\mu_{\nu_e}$ we find
\begin{equation}
\dot N_{\nu_\tau}=\int_{\mu_{\nu_\tau}}^{\mu_{\nu_e}}\!\!dE
\biggl[\frac{\Theta_{E}^2{\cal P}_{E} E^2}{2\pi^2}
+\int_{\mu_{\nu_\tau}}^{E}\!\!dE'
\sum_a{\cal W}^a_{EE'}
\frac{(g_\tau^a\Theta_{E}-g_e^a\Theta_{E'})^2 E^2E'^2}{4\pi^4}
\biggr]
\label{Zh}
\end{equation}
where we have restored an explicit superscript $a$ in the coupling
constants and a summation over targets. Note that for degenerate
neutrinos $N_{\nu}=\mu_{\nu}^3/6\pi^2$ whereas for electrons
$N_e=\mu_e^3/3\pi^2$ because of the additional r.h.\ spin degree of
freedom. Equation~(\ref{Zh}) is the starting point for our discussion
of flavor conversion in a SN core.

\newpage


\section{Flavor Conversion}
\subsection{Mixing of $\nu_e$ with $\nu_\mu$ or $\nu_\tau$}
We are now in a position to approach the problem of flavor conversion
between the degenerate $\nu_e$ and $\nu_\mu$ or $\nu_\tau$ in a SN core
quantitatively. To this end we consider two limiting cases for the
small mixing angle. For a $\nu_e$ near its Fermi surface with
$E=\mu_{\nu_e}$ we have in Eq.~(\ref{Xi}), taking $N=N_e$,
\begin{equation}
\frac{E}{E_r}=\frac{(68\,{\rm keV})^2}{\Delta m^2}\,
Y_eY_{\nu_e}^{1/3}\rho_{14}^{4/3},
\label{Sb}
\end{equation}
where $\rho_{14}$ is the density in units of $10^{14}\,\rm g\,cm^{-3}$
and $Y_j$ gives the abundance of species $j$ relative to baryons. If
$\Delta m^2$ is so large that $E/E_r\ll1$ we may effectively use the
vacuum mixing angle. In the opposite limit we may ignore
$\cos2\Theta_0$ in the denominator of Eq.~(\ref{Xi}). Hence we use
\begin{equation}
\vert\Theta_E\vert=\Theta_{0}\times\cases{1& ``vacuum''\cr
E_r/E&``medium''\cr}
\label{Sc}
\end{equation}
as our limiting cases.

Given enough time, the $\nu_\tau$ will reach the same chemical
potential as the $\nu_e$. Therefore, it will be most practical to
discuss the approach to chemical equilibrium in terms of a
dimensionless $\nu_\tau$ density
\begin{equation}
\eta\equiv N_{\nu_\tau}/N_{\nu_e}.
\label{Scc}
\end{equation}
Therefore, the equation of motion will be of the general form
\begin{equation}
\dot\eta=\Theta_{0}^2\times
\cases{\displaystyle{(9\pi)^{1/3}G_{\rm F}^2 N_e^{5/3}
\sum_a F_a^{\rm V}(\eta)}&``vacuum'',\cr
\displaystyle{\frac{(\Delta m^2)^2}{2\pi N_e}\,
\sum_a F_a^{\rm M}(\eta)}&``medium''.\cr}
\label{Sccc}
\end{equation}
For later convenience we have included the electron density $N_e$
rather than the baryon density as well as some numerical factors in the
overall normalization. The $F_a^{\rm V,M}$ are dimensionless functions
of $\eta$.
It is interesting that in the ``medium case'' the dimensionfull factors
do not involve Fermi's constant; a factor $G_{\rm F}^2$ from the
scattering rate cancels by $G_{\rm F}^{-2}$ from $\Theta_E^2$.

Beginning with the first term in Eq.~(\ref{Zh}) we use the neutrino
production rate ${\cal P}_E$ due to the reaction
$p+e\to n+\nu_e$ that was given in Eq.~(\ref{Ba}). An explicit
integration yields for the two limiting mixing angle cases defined in
Eq.~(\ref{Sc})
\begin{eqnarray}
F_{\rm CC}^{\rm V}(\eta)
&=&\frac{12}{5}
\left(\frac{\mu_{\nu_e}}{\mu_e}\right)^2
(1-\eta^{5/3}),
\nonumber\\
F_{\rm CC}^{\rm M}(\eta)&=&1-\eta
\label{Sd}
\end{eqnarray}
for the r.h.s.\ of Eq.~(\ref{Sccc}). For comparison with
Ref.~\cite{Turner} we note that in the vacuum limit, and at $t=0$ when
$\eta=0$, our result can be written as
$\dot N_{\nu_\tau}=\Theta_0^2\,(4G_{\rm
F}^2/10\pi^3)N_p\mu_{\nu_e}^5$ if we use $C_V^2+3C_A^2=4$.

Turning to elastic neutrino scatterings, i.e., to the term involving
${\cal W}^a_{EE'}$ in Eq.~(\ref{Zh}), we first consider nucleons as
targets, $\nu+N\to N+\nu$. This neutral-current (NC) reaction involves
identical amplitudes for all flavors whence our matrices $G$ can be
taken to be the $2{\times}2$ unit matrices for both proton and neutron
targets. Then Eq.~(\ref{Zh}) implies immediately that no flavor
conversion occurs in the limit of vacuum mixing where
$\Theta_E=\Theta_{E'}=\Theta_0$; hence $F_{\rm NC}^{\rm V}(\eta)=0$.
In the limit of medium mixing where $\Theta_E\propto E^{-1}$, however,
NC reactions do contribute to flavor conversion, contrary to statements
found in the literature \cite{Maalampi}, an issue discussed in
Appendix~A. By means of the function
${\cal W}^{p+n}_{EE'}$ given in Eq.~(\ref{Bd}) we find explicitly
\begin{equation}
F_{\rm NC}^{\rm M}(\eta)=\frac{1}{5}
\left(\frac{\mu_{\nu_e}}{m_N}\right)^2
\frac{1}{Y_e}
\sum_{j=p,n} X_j\frac{(C_V^2+7C_A^2)_j}{8}
\Bigl(1-\eta^{5/3}\Bigr)
\label{SSd}
\end{equation}
where $X_{p,n}$ are the density fractions of protons and neutrons with
$X_p+X_n=1$.
The flavor conversion rate is much smaller than that of CC scattering
Eq.~(\ref{Sd}) because it is suppressed by the ``recoil factor''
$(\mu_{\nu_e}/m_N)^2$. Put another way, damping is small because the
neutrino energy in a collision does not change very much so that even
in the medium $\Theta_E\approx\Theta_{E'}$. It is nevertheless
conceptually interesting that this process alone would suffice to reach
chemical equilibrium.

Next, we consider r.h.\ electrons as targets.  According to
Eq.~(\ref{Ub}) we have
$g_e^{\rm R}=g_\tau^{\rm R}\approx{1\over2}$ so that no flavor
conversion occurs in the vacuum case:
$F_{\rm R}^{\rm V}(\eta)=0$. For the medium case we find with the
approximate transition rates of Eq.~(\ref{Um})
\begin{equation}
F_{\rm R}^{\rm M}(\eta)=\frac{3}{32}\,
\frac{\mu_{\nu_e}}{\mu_e}\,
\Bigl({\textstyle{1\over80} - {1\over4}\eta^{2/3} + \eta
      + {1\over4}[\log(\eta)-{9\over4}]\eta^{4/3} -{1\over5}\eta^{5/3}}
\Bigr).
\label{UUa}
\end{equation}
Chemical equilibrium corresponds to $\eta=1$; it is easy to check that
$F_{\rm R}^{\rm M}(1)=0$.

For l.h.\ electrons we have according to Eq.~(\ref{Ub})
$g_e^{\rm L}\approx3/2$ and $g_\tau^{\rm L}\approx-{1\over2}$.
Therefore, we need to consider the vacuum as well as the medium case.
We find with Eq.~(\ref{Um})
\begin{eqnarray}
F_{\rm L}^{\rm V}(\eta)&=&\frac{1}{2}
\left(\frac{\mu_{\nu_e}}{\mu_e}\right)^3
\left(1-\eta^{1/3}\right)^3
\left({\textstyle 1 + {3\over5}\eta^{1/3} + {3\over10}\eta^{2/3} +
{1\over10}\eta}\right)
,\nonumber\\
F_{\rm L}^{\rm M}(\eta)&=&\frac{3}{32}\,
\frac{\mu_{\nu_e}}{\mu_e}\,
\left({\textstyle{9\over5}\eta^{-1/3}
+ {1\over4}\log(\eta) -{55\over16}
   + {5\over3}\eta^{1/3} + {1\over4}\eta^{2/3} -
     {67\over240}\eta^{4/3}}\right).
\label{UUb}
\end{eqnarray}
In the medium case the r.h.s.\ of the equation of motion diverges
initially when $\eta=0$. This singularity is unphysical because it is
related to the mixing angle $\Theta_E\propto E^{-1}$ becoming
``infinite'' at low neutrino energies; in reality $\vert\Theta_E\vert$
never exceeds $\pi/4$.

In the ``medium case'' the contribution of r.h.\ electrons is much
smaller than that of l.h.\ ones, in part because the function
${\cal W}_{EE'}$ is much smaller for r.h.\ states
(see Fig.~\ref{FigA}), and in part because for r.h.\ electrons the
integrand $(g_\tau\Theta_E-g_e\Theta_{E'})^2\propto(E-E')^2/E^2E'^2$ is
much smaller than for l.h.\ states where we have
$(g_\tau\Theta_E-g_e\Theta_{E'})^2\propto(3E+E')^2/E^2E'^2$.
In any event, we may now add the two contribution and obtain
\begin{equation}
F_{\rm R+L}^{\rm M}(\eta)=\frac{3}{32}\,
\frac{\mu_{\nu_e}}{\mu_e}\,
\left({\textstyle{9\over5}\eta^{-1/3}-{137\over40}
+{1\over4}(1+\eta^{4/3})\log(\eta) + {5\over3}\eta^{1/3} + \eta
- {101\over120}\eta^{4/3}-{1\over5}\eta^{5/3}}\right)
\label{UUc}
\end{equation}
for the total electron contribution.

We now collect our results. In the vacuum limit the only contributions
are from CC nucleon scattering, see Eq.~(\ref{Sd}), and from the
scattering on l.h.\ electrons, see Eq.~(\ref{UUb}). The electron
contribution is suppressed by a relative factor
${5\over24}(\mu_{\nu_e}/\mu_e)$, and it has a different structure of
the $\eta$ dependent term. In Fig.~\ref{FigB} (upper panel) we show
these terms; the electron contribution drops to zero much faster than
the nucleon term for $\eta\to1$. This is due to degeneracy effects
because as the $\nu_\tau$ Fermi sea fills up the accessible energy
difference $E-E'$ between initial- and final-state neutrinos decreases
so that an ever smaller fraction of the electrons is available as
scattering targets. For all practical purposes we may neglect the
electron term entirely so that the deviation from flavor equilibrium
relaxes as
\begin{equation}
\frac{d}{dt}\,(1-\eta)=-\frac{(1-\eta^{5/3})}{\tau}
\label{SumA}
\end{equation}
where
\begin{equation}
{\tau}{\Theta_0^2}=
\frac{5}{12(9\pi)^{1/3}G_{\rm F}^2 N_e^{5/3}}
\left(\frac{\mu_e}{\mu_{\nu_e}}\right)^2
=2.4{\times}10^{-10}\,{\rm s}
\left(\frac{10^{14}\,\rm g/cm^3}{Y_e\rho}\right)^{5/3}
\left(\frac{\mu_e}{\mu_{\nu_e}}\right)^2.
\label{SumAA}
\end{equation}

In the medium limit we have a contribution to the damping rate from CC
scattering on nucleons, given by Eq.~(\ref{Sd}). NC scattering also
contributes, see Eq.~(\ref{SSd}), but this term is suppressed by a
small ``recoil factor'' so that we may ignore it. NC and CC current
scattering on electrons both contribute; the combined result was given
in Eq.~(\ref{UUc}). However, compared with the nucleon result it is
suppressed by a factor ${3\over32}(\mu_{\nu_e}/\mu_e)$, while the
$\eta$ dependent terms are shown in Fig.~\ref{FigB} (lower panel).
Again, the electron contribution quickly drops to zero as flavor
equilibrium is approached. Thus, while electrons initially contribute
to flavor conversion, they may be neglected for the overall time scale
and we may use,
\begin{equation}
\frac{d}{dt}\,(1-\eta)=-\frac{(1-\eta)}{\tau}\label{SumB}
\end{equation}
where
\begin{equation}
{\tau}{\Theta_0^2}=
\frac{2\pi N_e}{(\Delta m^2)^2}
=1.8{\times}10^{-3}\,{\rm s}
\left(\frac{1\,{\rm keV}}{\sqrt{\Delta m^2}}\right)^4
\frac{Y_e\rho}{10^{14}\,\rm g/cm^3}\,.
\label{SumBB}
\end{equation}

In Fig.~\ref{FigC} we show contours of $\log_{10}(\tau\Theta_0^2)$
according to these results.  Considering the allowed range of vacuum
mixing angles it is evident that neutrino mass differences well in
excess of the cosmological bound of around $100\,{\rm eV}$ are required
to achieve flavor equilibrium on a time scale of seconds or faster.

\subsection{Sterile Neutrinos}
As another application of our equations we may easily calculate the
emission of ``sterile neutrinos'' $\nu_x$ from a SN core if these
particles mix with $\nu_e$, a problem previously discussed in
Ref.~\cite{Kainulainen}. The $\nu_x$ would escape directly from the
inner SN core so that the relevant figure of merit is the SN
energy-loss rate $\dot Q$, and the loss rate of lepton number $\dot
N_L$ into this channel. A general expression for $\dot N_L$ was given
in Eq.~(\ref{Ycccc}).

We consider, again, the two limiting cases for small mixing angles
defined in Eq.~(\ref{Sc}) and begin with the ``vacuum limit'' where
$\Delta m^2$ is so large that typically $E\ll E_r$. Assuming a small
vacuum mixing angle we may thus use $s_{\bf p}^2=4\Theta_0$ in
Eq.~(\ref{Ycccc}). We first consider the CC process $e+p\to n+\nu$ and
find with Eq.~(\ref{Ba})
\begin{eqnarray}
&\dot N_L&=\Theta_0^2 \frac{2}{5\pi^3}\,
G_{\rm F}^2 N_e\mu_{\nu_e}^5
\nonumber\\
&\dot Q&=\Theta_0^2 \frac{1}{3\pi^3}\,
G_{\rm F}^2 N_e\mu_{\nu_e}^6
\label{Oc}
\end{eqnarray}
where we have used $C_V^2+3C_A^2\approx4$. Similarly, we consider the
NC scattering process $\nu+N\to N+\nu$ which now does not involve any
destructive interference because $g_x=0$ for all processes. Therefore,
we obtain the same result as in Eq.~(\ref{Oc}) if we incorporate an
extra factor $(N_p+N_n)/N_e=1/Y_e$, and a further factor $1\over4$ for
the reduced coupling strength of NCs relative to CCs. (For NC
scattering we also take $C_V^2+3C_A^2\approx4$ for both protons and
neutrons.)
For electrons we use the approximate transition probabilities given in
Eq.~(\ref{Um}) and add the contributions from r.h.\ and l.h.\ electrons
from the start, recalling that $(g_e^{\rm R})^2 ={1\over4}$ while
$(g_e^{\rm L})^2 ={9\over4}$. Then we find the same results as
Eq.~(\ref{Oc}) except for an extra factor
${91\over768}(\mu_{\nu_e}/\mu_e)$ for $\dot N_L$ and
${23\over560}(\mu_{\nu_e}/\mu_e)$ for $\dot Q$. Clearly, we may neglect
the contribution of electrons relative to those of nucleons.
Therefore, lepton number is lost at a rate
\begin{equation}
\dot Y_L=-\Theta_0^2 G_{\rm F}^2 N_B^{5/3}\,
\frac{12(36\pi)^{1/3}}{5}\,(Y_e+{\textstyle{1\over4}})Y_{\nu_e}^{5/3}.
\label{Od}
\end{equation}
Because $Y_L\approx Y_e$ and $Y_{\nu_e}$ is $Y_e$ up to a factor of
order unity, we have
\begin{equation}
\frac{\dot Y_L}{Y_L^{8/3}}\approx
-\Theta_0^2G_{\rm F}^2 N_B^{5/3}
=-\Theta_0^2\,2.7{\times}10^{10}\,{\rm s^{-1}}
\frac{\rho}{10^{15}\rm\,g/cm^3}.
\label{Oe}
\end{equation}
Since most of the trapped energy is stored in the degenerate leptons,
energy is lost on a similar time scale. In order to avoid conflict with
standard SN physics, and especially with the observed duration of the
neutrino signal from SN 1987A, we must demand
$\Theta_0\alt 10^{-5}$, a bound in agreement with that found in
Ref.~\cite{Kainulainen}.

Next, we turn to the ``medium limit'' defined in Eq.~(\ref{Sc}) where
$\Delta m^2$ is small enough that typically $E\gg E_r$. For degenerate
$\nu_e$ this means $\mu_e\gg E_r$ or
$(\Delta m^2)^{1/2}\alt100\,\rm keV$ for the inner core of a SN.
It is easy to evaluate $\dot Q$ and $\dot N_L$ for the CC and NC
processes involving nucleons. However, in the present situation the
scattering term on electrons dominates because $\dot Q$ and $\dot N_L$
diverge if we use the simple approximation $\Theta_E=\Theta_0\,E_r/E$
of Eq.~(\ref{Sc}). We encountered this divergence before in
Eq.~(\ref{UUb}), but there we were able to ignore it because it
affected only the initial process of flavor conversion while the
overall time scale was dominated by the CC nucleon scattering process.
In the present case, the $\nu_x$ Fermi sea never fills up because they
continuously escape so that the resonance region with $E'=E_r$ always
remains within the region of integration.

In order to obtain a meaningful result we need to keep the full
expression~(\ref{Xiii}) for $s^2_{\bf p}$. However, this means that the
range of integration includes a certain regime where the mixing angle
in the medium is so large that even the ``sterile'' neutrinos are
trapped.  There will be a critical mixing angle $\Theta_c$ below which
the sterile neutrinos escape freely, and above which they are
essentially trapped. We assume that $\Theta_0\ll\Theta_c\ll 1$ so that
finally
\begin{equation}
\dot N_L=\Theta_0^2\frac{G_{\rm F}^2 N_e}{16\pi^3\mu_e} E_r^2
\int_{E_c}^{\mu_{\nu_e}}\!\! dE\int_{E_c}^E\!\!dE'
\frac{(E-E')(9E^3+E'^3)}{(E'-E_r)^2}
\label{OOa}
\end{equation}
where $E_c\equiv E_r(1+\Theta_0/\Theta_c)$. An explicit integration
yields, keeping only the lowest powers of $\Theta_0$ and of
$E_r/\mu_{\nu_e}$,
\begin{equation}
\dot N_L=\Theta_0\Theta_c
\,\frac{9}{80\pi^3}\,
\frac{G_{\rm F}^2N_e\mu_{\nu_e}^5E_r}{\mu_e}.
\label{OObb}
\end{equation}
Because of the resonance, the emission rate is proportional to
$\Theta_0\Theta_c$ rather than to $\Theta_0^2$.

The energy-loss rate is found by including an extra factor of $E'$ in
the integral Eq.~(\ref{OOa}). To lowest order we find $\dot Q=\dot N_L
E_r$ because of the resonance. Put another way, in the present
situation mostly neutrinos are emitted with energies around
$E_r\ll \mu_{\nu_e}$.  Hence lepton number is lost much faster than
energy. After the lepton number of the SN has been depleted by this
mechanism, assuming it dominates the standard diffusion transport, most
of the energy will still be trapped and has to be emitted by diffusion
to the surface. Therefore, a constraint on $\Delta m^2$ and $\Theta_0$
would require a much more detailed consideration of the cooling history
of a SN core with an early loss of lepton number. In this regard we
disagree with the conclusions of Ref.~\cite{Kainulainen} where even in
this case a bound was derived on the basis of a simple energy-loss
argument. There, the scattering on electrons and thus the quick
deleptonization was not included.

\newpage


\section{Conclusions}
Starting with the general and rigorous ``non-abelian collision
integral'' for mixed neutrinos of Ref.~\cite{RaffeltSS} we have derived
a simple and transparent kinetic equation for the occupation numbers of
two mixed neutrino flavors under the assumption of fast oscillations
relative to the collision rate, an approximation appropriate for a SN
core (``weak damping limit''). This form of the equation is both
suitable for practical calculations concerning neutrino flavor
conversion in a SN core, and also allows one to develop a better
understanding of the interplay between collisions and oscillations
encapsuled in the relatively complicated expression of
Ref.~\cite{RaffeltSS}.

We explicitly derived a number of simple general results, some of which
are in conflict with statements found in the literature.  It was
thought, for example, that two neutrino flavors could not reach
chemical equilibrium with each other if for each target species they
had identical scattering amplitudes. In this situation it was supposed
that the medium could not ``measure'' the flavor composition of a mixed
state.  We found that this conclusion is correct only if the mixing
angle is independent of the neutrino energy.  As another example, we
found that the emission rate of
sterile neutrinos which mix with $\nu_e$ with an angle $\Theta$ are
emitted with a rate ${1\over2}\sin^2(2\Theta)$ times the production
rate of $\nu_e$'s if the medium is transparent to the $\nu_e$'s. On the
other hand, if the $\nu_e$'s are approximately in thermal equilibrium,
the relative rate of gain for the sterile states is only
${1\over4}\sin^2(2\Theta)$.

We used our collision integral to calculate the flavor relaxation time
$\tau$ between the degenerate $\nu_e$ and the non-degenerate $\nu_\mu$
or $\nu_\tau$ in a SN core. A contour plot of $\tau\Theta_0^2$ (vacuum
mixing angle $\Theta_0$) was shown in Fig.~\ref{FigC} as a function of
the neutrino squared mass difference $\Delta m^2$, and of the density.
Neutrinos obeying the cosmological mass bound can not achieve chemical
equilibrium on a time scale below that of SN cooling.

We also calculated the time scale for the emission of the stored energy
of a SN core if $\nu_e$ mixes with a sterile species $\nu_x$. If
$\sqrt{\Delta m^2}\agt100\,\rm keV$ the vacuum mixing angle was found
to be constrained by $\Theta_0\alt 10^{-5}$. If
$\sqrt{\Delta m^2}\ll 100\,\rm keV$ the ``resonance energy'' where
$\nu_e$ and $\nu_x$ become degenerate is below the $\nu_e$ Fermi
surface, leading to quick deleptonization by virtue of the
scattering process $\nu_e+e\to e+\nu_x$ where the final-state $\nu_x$
carries the resonance energy. In this case, no simple constraint on
$\Theta_0$ can be derived.

\acknowledgements
We thank Leo Stodolsky for many discussions of the problem of neutrino
oscillations and damping. This paper is based, in part, on work to be
submitted as a doctoral thesis by G.S.\ to the
Ludwig-Maximilians-Universit\"at, Munich.

\newpage


\appendix{Neutral-Current Scattering}
There is a limiting case of our kinetic equation that deserves some
special attention. Neutral-current (NC) neutrino interactions cause a
large contribution to the scattering rate in a SN core. Do they also
contribute to flavor conversion? In order to answer this question we
first note that the relevant $G^a$ matrix in Eq.~(\ref{Xb}) is the unit
matrix because the NC scattering amplitude is the same for all neutrino
flavors; the corresponding transition rate is denoted by
${\cal W}^{\rm NC}_{\bf pp'}$. Therefore, the collision integral is
\begin{equation}
\dot\rho_{\bf p}=i[\rho_{\bf p},\Omega_{\bf p}]+
\int\!d{\bf p'} \Bigl[{\cal W}^{\rm NC}_{\bf p'p}
(\rho_{\bf p'}-{\textstyle{1\over2}}\{\rho_{\bf p'},\rho_{\bf p}\})
-{\cal W}^{\rm NC}_{\bf pp'}
(\rho_{\bf p}-{\textstyle{1\over2}}\{\rho_{\bf p'},\rho_{\bf p}\})
\Bigr] ,
\label{Na}
\end{equation}
where we have ignored all other processes. For the global density
matrix $\rho\equiv\int\rho_{\bf p} d{\bf p}$ we thus find
\begin{equation}
\dot\rho=i\int\!d{\bf p} [\rho_{\bf p},\Omega_{\bf p}]
\label{Nb}
\end{equation}
while the collision term vanishes identically. Therefore, it appears
that NCs do not contribute to flavor conversion \cite{Maalampi}.

Of course, {\it some\/} flavor conversion takes place by virtue of the
oscillation term Eq.~(\ref{Nb}).  As a first case we assume that in the
medium the mixing angle is dominated by the vacuum value $\Theta_0$
which is the same for all modes $\bf p$.  Then the precession of the
polarization vector of all modes is around the same direction. If we
start with a pure ensemble of $\nu_e$ the final $\nu_\tau$ population
is ${1\over2}\sin^2(2\Theta_0)$ of the total neutrino density. This
result remains valid even if the neutrinos are scattered around between
different modes: there is no flavor conversion beyond that caused by
free oscillations.

This situation is better understood in the weak damping limit where
oscillations are fast compared with the rate of collisions.  Then, the
trivial free oscillation effect takes place ``instantaneously'' after
preparing the ensemble. The subsequent change of the total $\nu_\tau$
density is found by setting $g_e=g_\tau=1$ in Eq.~(\ref{Yd}),
\begin{eqnarray}
\dot N_{\nu_\tau}=\int\!d{\bf p}\int\!d{\bf p'}\,
{\cal W}^{\rm NC}_{\bf pp'}\Bigl\{
(s_{\bf p}^2-s_{\bf p'}^2)(1+t_{\bf p}t_{\bf p'})
[f^\tau_{\bf p}(1-f^\tau_{\bf p'})-f^e_{\bf p}(1-f^e_{\bf p'})]
\>\>\>&& \nonumber\\
+[s_{\bf p}^2+s_{\bf p'}^2-t_{\bf p}t_{\bf p'}
(c_{\bf p}^2+c_{\bf p'}^2)]
[f^e_{\bf p}(1-f^\tau_{\bf p'})-f^\tau_{\bf p}(1-f^e_{\bf p'})]
\Bigr\}.&&
\label{Nc}
\end{eqnarray}
It is evident that Eq.~(\ref{Nc}) vanishes identically if the mixing
angle does not depend on $\bf p$.

This picture changes dramatically, however, when the medium mixing
angle is a function of $\bf p$. In this case $\dot N_{\nu_e}$ and
$\dot N_{\nu_\tau}$ do not, in general, vanish. This behavior is
understood if we recall that in the weak damping limit the neutrino
states between collisions are most naturally described as ``propagation
eigenstates''.  As both flavor components scatter identically by NCs
one arrives at the naive conclusion that collisions cause no flavor
conversion because the coherence between the two components is not
broken. However, if the momentum state changes in a collision, the
mixing angle of the new state is different from the original one so
that the new propagation eigenstate is a different combination of
flavor eigenstates. The subsequent adjustment by the assumed fast
oscillations leads to a certain amount of flavor conversion in each
collision.

This behavior can be made more explicit in the limit of small mixing
angles where Eq.~(\ref{Nc}) becomes
\begin{eqnarray}
\dot N_{\nu_\tau}=\int\!d{\bf p}\int\!d{\bf p'}\,
{\cal W}^{\rm NC}_{\bf pp'}\Bigl\{
(\Theta_{\bf p}^2-\Theta_{\bf p'}^2)
[f^\tau_{\bf p}(1-f^\tau_{\bf p'})-f^e_{\bf p}(1-f^e_{\bf p'})]
\>\>\>&&\nonumber\\
+(\Theta_{\bf p}-\Theta_{\bf p'})^2
[f^e_{\bf p}(1-f^\tau_{\bf p'})-f^\tau_{\bf p}(1-f^e_{\bf p'})]
\Bigr\}.&&
\label{Nd}
\end{eqnarray}
Moreover, in the limit $\Theta\to0$ the distributions $f^\tau_{\bf p}$
and $f^e_{\bf p}$ are in kinetic equilibrium so that with a non-zero
but small $\Theta^2$ they will deviate from equilibrium only to this
order. Hence, we may use the detailed-balance condition
\begin{equation}
\int d{\bf p'}\,\left[{\cal W}_{\bf pp'}^{\rm NC}
f_{\bf p}(1-f_{\bf p'})
-{\cal W}_{\bf p'p}^{\rm NC}
f_{\bf p'}(1-f_{\bf p}) \right]={\cal O}(\Theta^2)
\end{equation}
for both $\nu_\tau$ and $\nu_e$. Next, we multiply this equation with
$\Theta_{\bf p}^2$, apply $\int d{\bf p}$, and substitute
${\bf p}\leftrightarrow {\bf p'}$ in the second term under the
integrals. Therefore, the first line in Eq.~(\ref{Nd}) is of order
$\Theta^4$ and may thus be neglected:
\begin{equation}
\dot N_{\nu_\tau}=\int\!d{\bf p}\int\!d{\bf p'}\,
{\cal W}^{\rm NC}_{\bf pp'}
(\Theta_{\bf p}-\Theta_{\bf p'})^2
[f^e_{\bf p}(1-f^\tau_{\bf p'})-f^\tau_{\bf p}(1-f^e_{\bf p'})].
\label{Ne}
\end{equation}
If the ensemble consisted originally of $\nu_e$ only, and if the
distributions are in kinetic equilibrium to order $\Theta^2$, then
$f^e_{\bf p}>f^\tau_{\bf p}+{\cal O}(\Theta^2)$ until chemical
equilibrium is reached. Hence the r.h.s.\ of Eq.~(\ref{Ne}) is always
positive until the process of flavor conversion is complete.

Therefore, contrary to the naive conclusion stated above, flavor
equilibrium will be achieved by NC scatterings alone as long as
$\Theta_{\bf p}\not=\Theta_{\bf p'}$ for
$\vert{\bf p}\vert\not=\vert{\bf p'}\vert$.
This result, too, must be interpreted with care, however. If NC
interactions were the {\it only\/} interactions there would be no
difference between the refractive indices of the two flavors.
Then, the mixing angle in the medium would be strictly identical with
that in vacuum and thus independent of ${\bf p}$; hence flavor
equilibrium can not be achieved.

Thus, our result really means that CC interactions have a two-fold
effect on the process of flavor conversion. They cause different
scattering amplitudes for the two mixing flavors and thereby break the
coherence between the components of a mixed state, leading to
decoherence. Also, they cause different forward scattering amplitudes
and hence lead to a medium-induced component of the oscillation term.
This implies that the mixing angle in the medium becomes a function of
the neutrino energy, allowing NC interactions to contribute to flavor
conversion.

Of course, in a SN core the dominant NC scattering rate is on nucleons
which, because of their relatively large mass, do not allow for a large
amount of energy transfer in a collision with a neutrino.  In this case
the neutrino energy before and after a collision remains approximately
the same so that $\Theta_{\bf p}\approx\Theta_{\bf p'}$.  In this case
the impact of NCs on flavor conversion is relatively small as we saw
quantitatively in Sect.~III.


\appendix{Neutrino Interaction Rates}
In order to evaluate the equations of motion for the neutrino
distributions in a SN core we need the production rate ${\cal P}_E$ of
electron neutrinos with energy $E$, and the transition probability
${\cal W}_{EE'}^a$ for the scattering on a target species $a$ of a
neutrino with energy $E$ into a state with energy $E'$.

\subsection{Charged-Current Scatterings on Nucleons}
Beginning with ${\cal P}_E$, electron neutrinos can be produced and
absorbed by the charged-current (CC) reaction $p+e\leftrightarrow
n+\nu_e$. Because the effect of flavor conversion causes only a small
perturbation of the electron distribution we compute ${\cal P}_E$ from
the medium equilibrium properties. Moreover, we work in the limit of
highly degenerate leptons so that the occupation numbers of the
electrons are taken to be 1 below their Fermi surface, and 0 above. The
nucleons, however, are only partially degenerate; as a first
approximation we ignore nucleon Pauli blocking factors, and we ignore
correlation effects in the medium. Finally, the nucleons are
non-relativistic so that the CC scattering of an electron of energy $E$
produces a neutrino of the same energy.  Therefore, we approximate
${\cal P}_E=\sigma_E N_p$ where $N_p$ is the proton density and
$\sigma_E$ is the CC scattering cross section for electrons of energy
$E$; it is $\sigma_E=(C_V^2+3C_A^2) G_{\rm F}^2 E^2/\pi$ where $C_V$
and $C_A$ are the usual vector and axial weak couplings which in vacuum
are 1 and approximately 1.26, respectively. The exact value of $C_A$ in
a nuclear medium is not known, but likely it is suppressed somewhat so
that 1.0 is a more realistic value \cite{RaffeltS}.  Therefore, we use
$(C_V^2+3C_A^2)\approx4$ and find altogether \begin{equation}
{\cal P}_E\approx\frac{4 G_{\rm F}^2}{\pi}\,E^2N_p,
\label{Ba}
\end{equation}
accurate up to a factor of order unity. (Pauli blocking of nucleons can
not be neglected entirely; a suppression by up to a factor of 3 or so
occurs.)

\subsection{Neutral-Current Scatterings on Nucleons}
Turning to elastic neutrino scatterings we first consider nucleons as
targets, $\nu+N\to N+\nu$. Because the nucleons are non-relativistic we
may use
\begin{equation}
{\cal W}_{EE'}^{p,n}=N_{p,n}\frac{2\pi^2}{E'^2}\,
\frac{d\sigma_{p,n}}{dE'}
\label{Bb}
\end{equation}
where $\sigma_{p,n}$ is the scattering cross section on $p$ or $n$ for
neutrinos of energy $E$. A standard calculation yields \cite{Commins}
\begin{equation}
\frac{d\sigma}{dE'}=\frac{G_{\rm F}^2 m_N}{2\pi}
\Biggl[A+B\,\left(\frac{E'}{E}\right)^2+C\,\frac{E-E'}{E^2}\,m_N
\Biggr]
\label{Bc}
\end{equation}
where $A={1\over4}(C_A+C_V)^2$, $B={1\over4}(C_A-C_V)^2$, and
$C={1\over4}(C_A^2-C_V^2)$. In vacuum, for protons $C_V=1-
4\sin^2\Theta_{\rm W}$ and $C_A=1.26$ while for neutrons $C_V=-1$ and
$C_A=-1.26$. In a nuclear medium the axial charges for NCs are likely
suppressed by a similar amount as for CCs \cite{RaffeltS}.

Because the nucleons are relatively heavy compared with the neutrino
energies we go only to lowest order in $E/m_N$ where $m_N$ is the
nucleon mass.  In this approximation we find
\begin{equation}
{\cal W}^{p+n}_{EE'}=
\frac{\pi G_{\rm F}^2 m_N}{4 E^2}
\sum_{j=p,n}N_j\left[2(C_A^2+C_V^2)_j
+(C_A^2-C_V^2)_j\,\frac{E-E'}{E^2}\,m_N\right].
\label{Bd}
\end{equation}
To this order the maximum energy transfer in a collision is $2E^2/m_N$
so that ${\cal W}_{EE'}=0$ for $E'>E$ and for $E'<E(1-2E/m_N)$.

\subsection{Neutrino Electron Scattering}
The scattering of $\nu_e$ and $\nu_\tau$ on electrons can each be
described by an effective NC Hamiltonian. Taking the neutrino field
$\psi_\nu$ to represent a two-spinor in flavor space we have
$H_{e\nu}^{\rm int}=H_{e\nu}^{\rm L}+H_{e\nu}^{\rm R}$ with
\begin{equation}
H^{\rm L,R}_{e\nu}=\frac{G_{\rm F}}{2\sqrt2}\,
\overline\psi_e\gamma_\mu(1\mp\gamma_5)\psi_e\,
\overline\psi_\nu\gamma^\mu(1-\gamma_5)G^{\rm L,R}\psi_\nu.
\label{Ua}
\end{equation}
The upper sign refers to L (left-handed electrons), the lower to R
(right-handed electrons), and
\begin{eqnarray}
G^{\rm L}&=&\pmatrix{2\sin^2\Theta_{\rm W}+1&0\cr
                     0&2\sin^2\Theta_{\rm W}-1\cr},\nonumber\\
G^{\rm R}&=&\pmatrix{2\sin^2\Theta_{\rm W}&0\cr
                     0&2\sin^2\Theta_{\rm W}\cr}
\label{Ub}
\end{eqnarray}
where $\Theta_{\rm W}$ is the weak mixing angle. We will always use the
approximate value $\sin^2\Theta_{\rm W}\approx1/4$.

Because the electrons are ultra-relativistic the chirality states
coincide essentially with the left- and right-handed helicity states.
Therefore, the two electron helicities appear as two separate target
species. Treating the electrons as massless free Dirac particles we
find
\begin{equation}
{\cal W}_{EE'}^{\rm L,R}=
\int\!\frac{d\Omega_{\bf p}}{4\pi}\!
\int\!\frac{d\Omega_{\bf p'}}{4\pi}\!
\int\!\frac{d^3{\bf k}}{(2\pi)^3}\!
\int\!\frac{d^3{\bf k'}}{(2\pi)^3}
\frac{\vert{\cal M}^{\rm L,R}\vert^2}{2E2E'2\omega2\omega'}\,
n_{\bf k}(1-n_{\bf k'})\,(2\pi)^4\delta^4(p+k-p-k')
\label{Uc}
\end{equation}
where $p$ and $p'$ refer to the neutrinos, $k$ and $k'$ to the
electrons, and $n_{\bf k}$ are the electron occupation numbers.
The squared matrix elements are
$\vert{\cal M}^{\rm L}\vert^2=8G_{\rm F}^2(pk)(p'k')$ and
$\vert{\cal M}^{\rm R}\vert^2=8G_{\rm F}^2(pk')(p'k)$, respectively
\cite{Gandhi}. Performing all angular integrals explicitly we obtain
\begin{eqnarray}
{\cal W}^{\rm L}_{EE'}&=&\frac{G_{\rm F}^2}{30\pi}\,
\frac{E^3}{E'^2}
\int_0^\infty\!\! d\omega\,n_\omega(1-n_{\omega'})\,
A^{\rm L}(\omega/E),
\nonumber\\
{\cal W}^{\rm R}_{EE'}&=&\frac{G_{\rm F}^2}{30\pi}\,
\frac{E'^3}{E^2}
\int_0^\infty\!\! d\omega\,n_\omega(1-n_{\omega'})\,
A^{\rm R}(\omega/E'),
\label{Uf}
\end{eqnarray}
where $n_\omega$ is the electron occupation number at energy $\omega$
(we have assumed isotropy of the medium) and $\omega'=\omega+E -E'$.
Moreover, we have
\begin{equation}
A^{\rm L,R}(\xi)=\cases{10\xi^3\pm5\xi^4+\xi^5&if $\xi\le1$,\cr
                        10\xi^2\pm5\xi+1&if $\xi\ge1$,\cr}
\label{Ug}
\end{equation}
where the upper sign refers to L, the lower sign to R.

We stress that ${\cal W}_{EE'}$ is different for l.h.\ and r.h.\
electron targets because the angular momentum budget of these reactions
is different. In the CM system $\nu$-$e_L$ scattering is isotropic,
while $\nu$-$e_R$ scattering has a $(1-\cos\theta)^2$ angular
distribution, i.e., it is forward peaked \cite{Commins}. This implies
that in the system of the background medium $\nu$-$e$ scattering is
more efficient at transferring energy to l.h.\ electrons than to r.h.\
ones.

In order to perform the $\omega$-integration we take the electrons to
be highly degenerate, i.e., we take $n_\omega=1$ for $\omega\le\mu_e$
and 0 otherwise. Moreover, we take the neutrino energies to be less
than $\mu_e$, and we note that neutrinos can only down-scatter; hence
we have $\mu_e>E\ge E'$. Thus, only electrons with energies above
$\mu_e-E+E'$ contribute and we have
\begin{equation}
\int_0^\infty d\omega\,n_\omega(1-n_{\omega'})\,\ldots
\longrightarrow
\int_{\mu_e-E+E'}^{\mu_e}d\omega\,\ldots\,.
\label{Uh}
\end{equation}
Then we find
\begin{eqnarray}
{\cal W}^{\rm R}_{EE'}&=&W_0\,
\frac{(E-E')E'}{E^2}\,
W^{\rm R}(E/\mu_e,E'/\mu_e) ,
\nonumber\\
\nonumber\\
{\cal W}^{\rm L}_{EE'}&=&W_0\times\cases{\displaystyle
\frac{(E-E')E}{E'^2}\,
W^{\rm L}_1(E/\mu_e,E'/\mu_e)
&if $2E\le\mu_e+E'$,\cr
\displaystyle\frac{\mu_e^4}{E^2E'^2}\,
W^{\rm L}_2(E/\mu_e,E'/\mu_e)
&if $2E>\mu_e+E'$,\cr}\label{Ui}
\end{eqnarray}
where $W_0\equiv\pi G_{\rm F}^2 N_e/\mu_e$ and
\begin{eqnarray}
W^{\rm R}(x,y)&=&{\textstyle{1\over60}}
[60-30(2x-y)+(20x^2-25xy+11y^2)],\nonumber\\
W^{\rm L}_1(x,y)&=&{\textstyle{1\over60}}
[60-30(x-2y)+(11x^2-25xy+20y^2)],\nonumber\\
W^{\rm L}_2(x,y)&=&{\textstyle{1\over60}}
[-1-6y-15y^2 +20(2x^3-y^3)-15(2x^4-4x^3y+y^4)\nonumber\\
&&+6(7x^5-15x^4y+10x^3y^2-y^5)\nonumber\\
&&-29x^6+36x^5y-45x^4y^2+20x^3y^3-y^6].
\nonumber
\end{eqnarray}

These expressions are already fairly complicated, and the number of
terms multiplies enormously once we perform the integrals over $dE'$
and $dE$ to obtain the damping rate. Therefore, we include only the
leading terms in $E/\mu_e$ and $E'/\mu_e$ since the neutrino energies
are always smaller than the electron chemical potential. To this order
we use
\begin{eqnarray}
{\cal W}^{\rm R}_{EE'}&=& W_0\frac{(E-E')E'}{E^2},\nonumber\\
{\cal W}^{\rm L}_{EE'}&=& W_0\frac{(E-E')E}{E'^2}.
\label{Um}
\end{eqnarray}
In Fig.~\ref{FigA} (upper panel) we show $({\cal W}^{\rm R}_{EE'}/W_0)
(E'/E)^2$ for $E={1\over4}\mu_e$, ${1\over2}\mu_e$, ${3\over4}\mu_e$,
and $\mu_e$ as a function of $E'/\mu_e$. The solid lines refer to the
exact result Eq.~(\ref{Ui}) while the dotted lines refer to the
lowest-order result Eq.~(\ref{Um}). In the lower panel we show the same
graphs for ${\cal W}^{\rm L}_{EE'}$.



\figure{The $\eta$-dependent part of the functions $F_a^{\rm V}(\eta)$
(upper panel) and $F_a^{\rm M}(\eta)$ (lower panel). The curves from
CC scattering on nucleons are according to Eq.~(\ref{Sd}), while for
electron scattering the ``vacuum'' curve is from Eq.~(\ref{UUb}), the
``medium'' curve from Eq.~(\ref{UUc}).\label{FigB}}

\figure{Contour plot for $\tau\Theta_0^2$ according
to Eqs.~(\ref{SumAA}) and~(\ref{SumBB}), taking
$\mu_{\nu_e}/\mu_e\approx1$. The contours are marked with
$\log_{10}(\tau\Theta_0^2)$.\label{FigC}}

\figure{Transition probability ${\cal W}_{EE'}$ for
$\nu$-$e$-scattering for r.h.\ electrons (upper panel) and l.h.\
electrons (lower panel). Solid lines: Exact result according to
Eq.~(\ref{Ui}). Dotted lines: Lowest-order approximation according to
Eq.~(\ref{Um}).\label{FigA}}


\begin{references}
\bibitem{Maalampi}J. Maalampi and J. T. Peltoniemi,
    Phys. Lett. B {\bf 269}, 357 (1991).
\bibitem{Turner}M. Turner, Phys. Rev. D {\bf 45}, 1066 (1992).
\bibitem{Kainulainen}K.~Kainulainen, J.~Maalampi, and
    J.~T.~Peltoniemi, Nucl. Phys. B {\bf 358}, 435 (1991).
\bibitem{Stodolsky}L. Stodolsky, Phys. Rev. D {\bf 36}, 2273 (1987).
\bibitem{Flaig}J. Flaig, {\it Diplom}-Thesis, Univ.\ of Munich
    (1989), unpublished.
\bibitem{RaffeltSS}G. Raffelt, G. Sigl, and L. Stodolsky, Report,
    MPI-Ph/92-65 (1992).
\bibitem{Dolgov}A. Dolgov, Sov. J. Nucl. Phys. {\bf 33}, 700 (1981).
\bibitem{RaffeltS}G. Raffelt and D. Seckel, Report (1992).
\bibitem{Commins}E. D. Commnis and P. H. Bucksbaum, {\sl Weak
Interactions of Leptons and Quarks\/} (Cambridge University Press,
Cambridge, 1983).
\bibitem{Gandhi}{K. J. F. Gaemers, R. Gandhi, and J. M. Lattimer,
    Phys. Rev. D {\bf 40}, 309 (1989).}
\end{references}
\end{document}